\documentclass[pre,aps,twocolumn,superscriptaddress,notitlepage,nofootinbib]{revtex4-1}
\usepackage{amssymb,epsfig,amsmath,bm,subfigure,graphicx,textcomp,url,float,color,cases}

\usepackage{epsfig}
\usepackage{textcomp}
\usepackage[normalem]{ulem}

\usepackage[colorlinks=true, urlcolor=blue, anchorcolor=blue, citecolor=blue,filecolor=blue,linkcolor=blue,menucolor=blue]{hyperref}

\usepackage{color}

\usepackage[titletoc,title]{appendix}

\begin{document}

\title{Fluctuations of blowup time in a simple model of a super-Malthusian catastrophe}

\author{Baruch Meerson}
\affiliation{Racah Institute of Physics, Hebrew University of Jerusalem, Jerusalem 91904, Israel}

\begin{abstract}
Motivated by the paradigm of a super-Malthusian population catastrophe, we study a simple stochastic population model which exhibits a finite-time blowup of the population size and is strongly affected by intrinsic noise. We focus on the fluctuations of the blowup time $T$ in the asexual binary reproduction model $2A \to 3A$, where two identical individuals give birth to a third one. We determine exactly the average blowup time as well as the probability distribution $\mathcal{P}(T)$ of the blowup time and its moments. In particular, we show that the long-time tail $\mathcal{P}(T\to \infty)$ is purely exponential. The short-time tail $\mathcal{P}(T\to 0)$  exhibits an essential singularity at $T=0$, and it is dominated by a single (the most likely) population trajectory which we determine analytically.
\end{abstract}

\maketitle

\section{Introduction}
\label{intro}

It is argued that the human population of the Earth exhibits
a super-Malthusian (that is a faster-than-exponential) growth that should ultimately lead to a finite-time blowup \cite{Johanssen2001}. In general, a finite-time blowup in a population model can occur due to a positive feedback when the population growth rate increases faster than linearly with the population size. A simple example is provided by the following nonlinear ordinary differential equation (ODE) for the population size $n=n(t)$:
\begin{equation}
\label{DE}
\dot{n}(t) = \alpha n^2(t)\,,\quad \alpha >0\,,
\end{equation}
where the dot denotes the time derivative. The solution of this equation,
\begin{equation}
\label{determ}
n(t) = \frac{n_0}{1-\alpha n_0 t},
\end{equation}
where $n_0\equiv n(t=0)>0$ is the initial condition, blows up in a finite time $T=(\alpha n_0)^{-1}$.

In the presence of noise the blowup time becomes a random quantity, and it is interesting to determine its statistics. One way of addressing this class of problems is to interpret Eq.~(\ref{DE}) as the equation of motion of an overdamped particle, with the coordinate $n=n(t)$, in a repulsive potential $V(n) \sim - n^3$. By adding a  noise term to  Eq.~(\ref{DE}), one turns this mean-field equation into a Langevin equation:
\begin{equation}
\label{Langevin}
\dot{x}(t) = \alpha x^2(t)+ \eta(t)\,,
\end{equation}
where $\eta(t)$ is white Gaussian noise, and we have replaced $n$ by $x$. The finite-time blowup statistics for this model and its extensions has been recently studied in Ref. \cite{KM2025}, see also Refs.~\cite{Ryabov2016,Ornigotti,SilerPRL,Ryabov}.

Here we take a different approach by observing that Eq.~(\ref{DE}) also provides a mean-field description to the simple Markovian stochastic model of asexual binary reproduction $2A \to 3A$, where two identical $A$-individuals, which we will call particles, give birth to a third one. The master equation for this stochastic process is
\begin{equation}\label{master}
 \!\dot{P}_n(t)\!=\! \frac{\beta (n-1) (n-2)}{2} P_{n-1}(t) - \frac{\beta n (n-1)}{2} P_{n}(t),
\end{equation}
where $P_n(t)$ is the probability of observing $n$ particles at time $t$, and $\beta=2\alpha$.  We suppose that there are $m\geq 2$ particles at $t=0$, so the initial condition is
\begin{equation}\label{incond}
P_n(t=0) = \delta_{nm}\,.
\end{equation}
By rescaling time, $\beta t \to t$, we can get rid of the rate coefficient $\beta$. Therefore, from now on we set $\beta=1$.

In Sec.~\ref{stats} we determine exactly the average blowup time and the probability distribution of the blowup time for the $2A\to 3A$ model. We also determine several first moments of the distribution and provide an additional insight into the short-time tail of the probability distribution of the blowup time by employing the optimal fluctuation method (OFM). In Sec. \ref{conclusions} we briefly discuss our main results and compare them with those  for the Langevin equation~(\ref{Langevin}), obtained in Ref. \cite{KM2025}.

\section{Statistics of blowup time}
\label{stats}

\subsection{Average blowup time}
\label{theory}

As a warmup, let us calculate the average blowup time $\Theta(m)$ as a function of the initial number of particles $m$. This quantity can be determined from the  backward master equation, see, \textit{e.g.}, Ref. \cite{Gardiner}. For the process $2A \to 3A$ the backward master equation is
\begin{equation}
\label{avtimeeq}
r(m)\left[\Theta(m+1)-\Theta(m)\right]= -1,
\end{equation}
where $r(m) =m(m-1)/2$. The blowup,  which can be viewed as the first passage to infinity, is described by the absorbing boundary condition \cite{Gardiner}
\begin{equation}
\label{BCinfinity}
\Theta(m\to+\infty) = 0\,.
\end{equation}
The solution of the problem (\ref{avtimeeq}) and (\ref{BCinfinity}) is remarkably simple:
\begin{equation}
\label{T-int-m}
\Theta(m) =\frac{2}{m-1}\,.
\end{equation}
In particular, for $m=2$ (there are exactly two particles at $t=0$) we obtain $\Theta(2)\equiv \langle T\rangle =2$. It is twice as large as the mean-field prediction $t_0=1$ for the same initial number of particles. For $m\gg 1$ the mean-field prediction agrees with the exact result as to be expected.  Notice that the average blowup time (\ref{T-int-m}) falls off quite slowly as a function of $m$.

\subsection{Probability distribution of blowup time}
\label{distr}

A standard method \cite{Gardiner,SR} of determining the probability distribution $\mathfrak{p}(T,m)$ of the first passage time for a Markovian jump process which starts from an arbitrary $n=m$, operates with the Laplace transform of $\mathfrak{P}(T,m)$,
\begin{equation}
\label{LT}
\Pi(s,m)=\int_0^\infty dT\,e^{-sT} \mathfrak{p}(T,m)\,,
\end{equation}
where $s$ is a complex parameter. This Laplace transform obeys another recursive equation \cite{Gardiner} which, for
the jump process $2A \to 3A$, has the form
\begin{equation}\label{Pieq}
r(m)\left[\Pi(s,m+1)-\Pi(s,m)\right]=s \Pi(s,m)\,.
\end{equation}
The blowup is described by the boundary condition \cite{Gardiner}
\begin{equation}
\label{BC1}
\Pi(s,m\to+\infty)=1\,.
\end{equation}
The general solution of Eq.~(\ref{Pieq}) is
\begin{equation}\label{Pigen}
\Pi_{\text{gen}}(s,m) =C \, \frac{\mu(s)^+_{m-2} \,\mu(s)^-_{m-2}}{\Gamma (m-1) \Gamma (m)}\,,
\end{equation}
where
$\mu(s)^{\pm} =  (1/2)(3\pm \sqrt{1-8 s})$, $(z)_{k}\equiv\Gamma(z+k)/\Gamma(z)$ is the Pochhammer symbol, $\Gamma(\dots)$ is the gamma function, and $C$ is arbitrary constant. Taking the limit of $m\to \infty$ and using the boundary condition~(\ref{BC1}), we determine $C$ and arrive at the following result:
\begin{equation}\label{Pim}
\Pi(s,m) =\frac{\Gamma \left(m-\frac{1}{2}-\frac{1}{2} \sqrt{1-8 s}\right) \Gamma
   \left(m-\frac{1}{2}+\frac{1}{2} \sqrt{1-8 s}\right)}{\Gamma (m-1) \Gamma (m)}
\end{equation}
The inverse Laplace transform of this expression yields the blowup time distribution:
\begin{equation}
\label{inversetr}
\mathfrak{p}(T,m)= \frac{1}{2\pi i}\int_{\gamma-i \infty}^{\gamma+i\infty} e^{sT} \Pi(s,m) \,ds\,.
\end{equation}
Equations~(\ref{Pim}) and~(\ref{inversetr}) solve the problem of finding the probability distribution of blowup times for any $m$.  For concreteness, let us focus on $m=2$ where Eq.~(\ref{Pim}) simplifies to
\begin{equation}\label{Pi2}
\Pi(s,2)=\frac{2 \pi  s} {\cos \left(\frac{\pi}{2} \,\sqrt{1-8 s}\right)}\,.
\end{equation}
As is evident from this expression, the Laplace transform $\Pi(s,2)$ has infinitely many simple poles,
located on the negative real axis of $s$ at
$s\equiv s_k=-k(k+1)/2$, where $k=1, 2,\dots$. Near the $k$-th pole we have
\begin{equation}\label{poles}
\Pi(s,2)\simeq \frac{a_k}{s-s_k}\,.
\end{equation}
The inverse Laplace transform~(\ref{inversetr}) yields an exact representation of $\mathfrak{p}(T,2)\equiv \mathcal{P}(T)$ in terms of an infinite series,
\begin{equation}
\label{PTresult}
 \mathcal{P}(T)= \sum _{k=1}^{\infty } a_k e^{-\frac{1}{2} k (k+1) T}\,,
\end{equation}
where the amplitudes
\begin{equation}\label{a(k)}
  a_k=\frac{1}{2} (-1)^{k+1} k (k+1) (2 k+1)
\end{equation}
coincide with the amplitudes of the simple poles (\ref{poles}).

As a consistency check, let us reproduce the average value $\langle T\rangle$ of the blowup time $T$. We multiply Eq.~(\ref{PTresult}) by $T$, integrate it over $T$ term by term, and sum up the resulting number series:
\begin{equation}\label{Tav}
\langle T\rangle = \sum _{k=1}^{\infty }\frac{2 (-1)^{k+1} (2 k+1)}{k (k+1)} =2\,.
\end{equation}
As to be expected, this result coincides with our result $\Theta(2)=2$ in Eq.~(\ref{T-int-m}).
In the same straightforward way one can calculate the higher moments of the distribution,
\begin{equation}\label{Mp}
M_p=\int_0^{\infty} T^p \mathcal{P}(T)\,dT\,,\quad p=2,3,4,\dots\,,
\end{equation}
and obtain
\begin{eqnarray}
  M_2 &=& \frac{4}{3} \left(\pi^2-6\right),\\
  M_3 &=& 8 \left(12-\pi^2\right),\\
  M_4 &=& \frac{16}{15} \left(7\pi^4+120 \pi ^2-1800\right),\\
  M_5 &=& 32 \left(1680-7 \pi^4-100 \pi ^2\right), \\
  M_6 &=& \frac{64}{21} \left(31 \pi ^6+2646 \pi ^4+35280 \pi ^2-635040\right),
  \label{momentsT}
\end{eqnarray}
\textit{etc.}

\begin{figure}
\includegraphics[width=0.46\textwidth,clip=]{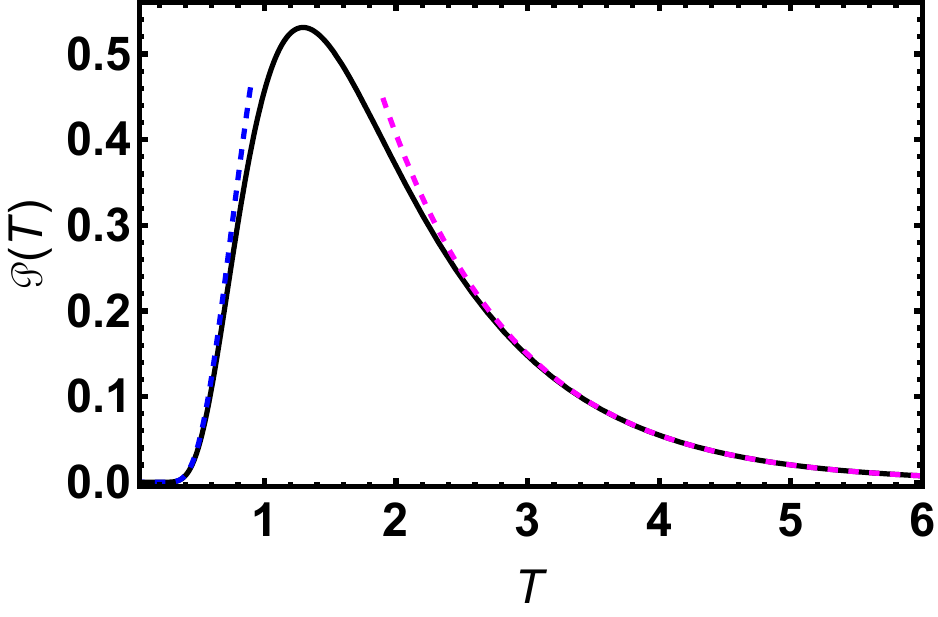}
\caption{The blowup time distribution $\mathcal{P}(T)$ for $m=2$, as described by Eqs.~(\ref{PTresult}) and~(\ref{a(k)}) (solid line). Dashed lines: the long-time asymptotic~(\ref{PlargeT}) and the short-time asymptotic~(\ref{PsmallT}).}
\label{P(T)fig}
\end{figure}

Figure \ref{P(T)fig} shows a plot of the blowup time probability distribution $\mathcal{P}(T)$ as described by Eqs.~(\ref{PTresult}) and~(\ref{a(k)}), alongside with the long-time and short-time tails of the distribution. The long-time tail,
\begin{equation}\label{PlargeT}
 \mathcal{P}(T\gg \langle T \rangle)\simeq 3 \,e^{-T}\,,
\end{equation}
which describes the probability distribution of unusually slow blowups, is determined by the slowest-decaying term $k=1$ of the series Eq.~(\ref{PTresult}). This term comes from the pole closest to the origin, $s=-1$, of the Laplace transform $\Pi(s,2)$.

There is also a simple physical picture behind the long-time tail $\mathcal{P}(T)\simeq 3\,e^{-T}$.  Indeed, consider an extreme situation when the two particles, introduced at $t=0$, do not reproduce even once during the whole time $T$. The probability cost of such an extreme event is exactly $e^{-T}$.

In the opposite limit of small $T$, that is for unusually fast blowups, all the terms in the series (\ref{PTresult}) contribute, which makes the series representation somewhat inconvenient. Here we utilize the fact that the $T\to 0$ asymptotic of $\mathcal{P}(T)$ is determined by the $s\to +\infty$ asymptotic of the Laplace transform $\Pi(s,2)$ in Eq.~(\ref{Pi2}). The latter has the form
\begin{equation}\label{larges}
 \Pi(s\to \infty,2) \simeq  4\pi s \,e^{-\pi \sqrt{2s}}\,.
\end{equation}
Its inverse Laplace transform, up to subleading corrections, is
\begin{equation}\label{PsmallT}
  \mathcal{P}(T\ll \langle T \rangle)  \simeq \frac{\sqrt{2}\, \pi ^{7/2}}{T^{7/2}}\, e^{-\frac{\pi ^2}{2 T}}\,.
\end{equation}
A salient feature of this tail is an essential singularity at $T=0$, which is also evident in Fig.~\ref{P(T)fig}.

In the next  subsection we will reproduce, up to the pre-exponential factor, the short-time tail~(\ref{PsmallT}) by employing the OFM. In this way we will also determine the optimal population trajectory, conditioned on an unusually fast blowup.

Now let us briefly return to $m>2$, where Eq.~(\ref{Pim}) yields
\begin{eqnarray}
  \Pi(s,3)&=&2 \pi  s (s+1) \sec \left(\frac{\pi}{2} \sqrt{1-8 s}\right)\,, \label{m3}\\
 \Pi(s,4)&=&\frac{2}{3} \pi  s (s+1) (s+3) \sec \left(\frac{\pi}{2} \sqrt{1-8 s}\right)\,, \label{m4}
\end{eqnarray}
\textit{etc.}
As one can see, the factors $(s+1)$, $(s+1)(s+3)$, $\dots$ in Eqs.~(\ref{m3}), (\ref{m4}), $\dots$, cancel one, two, $\dots$, consecutive poles $s_k=-k(k+1)/2$, starting from the pole $s_1=-1$. The long-time tails of the corresponding distributions $\mathfrak{p}(T,m)$ change accordingly: for $m=3, 4, \dots $ we obtain $\mathfrak{p}(T\to \infty,3) \sim e^{-3T}$, $\mathfrak{p}(T\to \infty,4) \sim e^{-6T}$, \textit{etc.} These long-time tails correspond to extreme situations when no reproduction events occur during nearly the whole time $T$ in the systems of $3$, $4$, \textit{etc.} particles.

\subsection{Short-time tail and optimal trajectory}
\label{sub:0}

\begin{figure}
\includegraphics[width=0.46\textwidth,clip=]{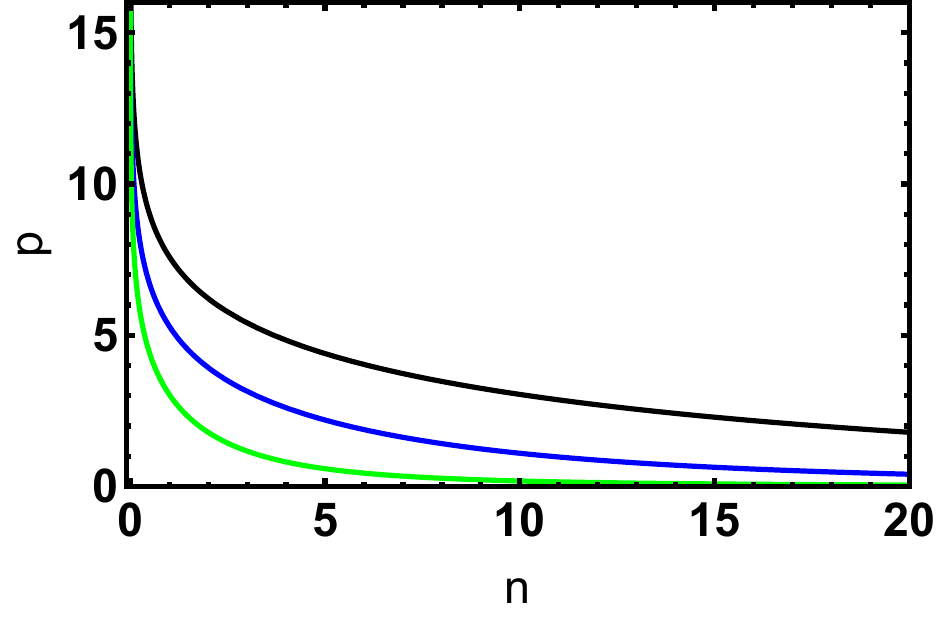}
\caption{The phase portrait of the system in the limit of $T\to 0$. Shown are three phase trajectories, as described by Eq.~(\ref{p(n)}) for $E=10^3$, $10^2$ and $10$ (from top to bottom). In view of Eq.~(\ref{TvsE}), the blowup times in these cases are $0.070\dots$, $0.222\dots$ and $0.702\dots$, respectively.}
\label{phaseplane}
\end{figure}

The short-time tail of $\mathcal{P}(T)$ can be described by the OFM. (The OFM and its analogs  are also known under the names of  weak-noise theory, dissipative WKB approximation, instanton method, eikonal approximation, geometrical optics, macroscopic fluctuation theory, \textit{etc.}).  The ``real space" OFM that we employ here involves applying a WKB-type ansatz $P_n(t) = \exp[-S(n,t)]$ to the master equation (\ref{master}), assuming a strong inequality $n\gg 1$ and treating $n$ as a continuous variable \cite{Kubo1973,Dykman1994,KesslerShnerb,MS2009,EscuderoKamenev,AM2010,AM2017}. [For $m=O(1)$, the assumption $n\gg 1$ is invalid at earlier times in this  model. This, however, introduces only a subleading-order error to the final result.]   In the leading order, one obtains an evolution equation for the function $S(n,t)$:
\begin{equation}\label{HJeq}
\partial_t S+ \frac{n^2}{2}\left(e^{\partial_n S}-1\right)=0\,.
\end{equation}
This equation has the form of a Hamilton-Jacobi equation of classical mechanics \cite{LLMechanics},  $\partial_t S+H(n,\partial_nS) = 0$. The effective Hamiltonian is
\begin{equation}\label{H}
H(n,p)= \frac{n^2}{2} \left(e^p-1\right)\,,
\end{equation}
where $n$ and $p$ are the canonically conjugated ``coordinate" and ``momentum".  The Hamilton equations have the form
\begin{eqnarray}
  \dot{n} &=& \frac{\partial H}{\partial p} = \frac{n^2}{2}\,e^p\,,\label{ndot}\\
  \dot{p} &=& -\frac{\partial H}{\partial n} = -n \left(e^p-1\right)\label{pdot}\,.
\end{eqnarray}
Since the Hamiltonian (\ref{H}) does not depend on time, it is a constant of motion, $H(n,p)=E =\text{const}$, which gives
\begin{equation}\label{p(n)}
p(n,E)=\ln \left(1+\frac{2E}{n^2}\right)\,.
\end{equation}
This equation with different values of the constant ``energy" $E$ describes the phase portrait of the system, see Fig.~\ref{phaseplane}. As we will see shortly, $E$ is determined by the blowup time $T$ that we condition the process on.  $E=0$ corresponds to the invariant manifold $p=0$, where Eq.~(\ref{pdot}) is obeyed automatically, and Eq.~(\ref{ndot}) coincides with the  mean-field equation~(\ref{DE}).

Plugging Eq.~(\ref{p(n)}) into Eq.~(\ref{ndot}), we obtain
\begin{equation}\label{dndt}
\dot{n}(t) = \frac{n^2}{2}+E\,.
\end{equation}
Integrating this simple equation from $n=0$ to infinity, we obtain the relation between the blowup time $T$ and the (\textit{a priori} unknown) ``energy" $E$:
\begin{equation}\label{TvsE}
T=\int_0^{\infty} \frac{dn}{\frac{n^2}{2}+E} = \frac{\pi}{\sqrt{2E}}\,.
\end{equation}
Therefore,
\begin{equation}\label{EvsT}
E=\frac{\pi^2}{2 T^2}\,.
\end{equation}
As one can see, the shorter is the blowup time $T$, the larger energy $E$ is required.
Now we can calculate the action along this trajectory:
\begin{eqnarray}
 S_0 &=& \int_0^{\infty} p(n,E) dn - ET \nonumber \\
 &=& \int_0^{\infty} \ln \left(1+\frac{2E}{n^2}\right) dn -ET\,.
 \label{S01}
\end{eqnarray}
Evaluating the integral and using Eq.~(\ref{EvsT}), we obtain
\begin{equation}\label{S02}
S_0 = \frac{\pi  \sqrt{E}}{\sqrt{2}} = \frac{\pi^2}{2T}\,,
\end{equation}
The probability distribution $\mathcal{P}(T)$ is given, in the leading order of the OFM,  by the relation
\begin{equation}\label{OFMprob}
\mathcal{P}(T) \sim e^{-S_0} = e^{-\frac{\pi^2}{2T}}\,.
\end{equation}
This result is asymptotically exact at $T \to 0$, when the action $S_0$ tends to infinity. As we can see, Eq.~(\ref{OFMprob})  agrees in the leading order  with the short-time tail (\ref{PsmallT}) that we obtained from the exact solution. It misses, however, the important pre-exponential factor $\sim T^{-7/2}\ll 1$ which appears in Eq.~(\ref{PsmallT}). The calculation of this factor in the OFM formalism would demand going to the subleading order, which we do not pursue here.

\begin{figure}
\includegraphics[width=7.89cm]{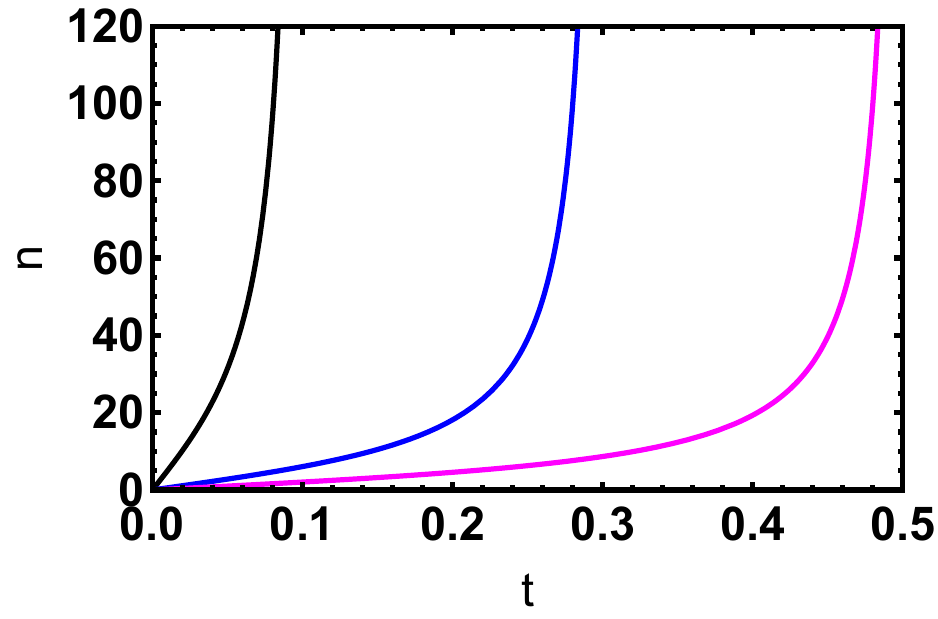}
\caption{The optimal population trajectory conditioned on unusually fast blowup times $T=0.1$, $0.3$ and $0.5$, as described by Eq.~(\ref{nvst}).}
\label{n(t)}
\end{figure}

Note that we succeeded in evaluating the short-time tail~(\ref{OFMprob}) without explicitly calculating the optimal population trajectory $n=n(t;T)$ which dominate this tail. The optimal trajectory, however, is interesting in its own right. To determine it, we return to Eq.~(\ref{dndt}) and
obtain
\begin{eqnarray}
  t = \int_0^{n} \frac{d\nu}{\frac{\nu^2}{2}+E} &=&  \sqrt{\frac{2}{E}}\,\arctan
\left(\frac{n}{\sqrt{2E}}\right) \nonumber\\
  &=& \frac{2 T}{\pi} \arctan\left(\frac{n T}{\pi }\right),
  \label{tvsn}
\end{eqnarray}
where we have again used Eq.~(\ref{EvsT}). Inverting Eq.~(\ref{tvsn}), we  obtain
\begin{equation}\label{nvst}
n(t;T) = \frac{\pi}{T}  \tan \left(\frac{\pi  t}{2 T}\right)\,, \quad t<T\,.
\end{equation}
Equation~(\ref{nvst}) describes the most likely trajectory of the population blowup conditioned on an unusually short blowup time $T$. Figure~\ref{n(t)} shows
these trajectories for $T=0.1$, $0.3$ and $0.5$. It is instructive to compare Eq.~(\ref{nvst}) with Eq.~(\ref{DE}) which  describes the finite-time blowup predicted by the mean-field theory.

\section{Discussion}
\label{conclusions}

We studied the blowup time statistics in a simple asexual binary reproduction model $2A \to 3A$ which features a super-Malthusian catastrophe. As we have seen, this model admits exact solutions for the complete probability distribution of the blowup time and its moments. We also extracted and interpreted the short- and long-time tails of the distribution. The short-time tail, which exhibits an essential singularity, is amenable to the OFM. In this way we also provided an additional insight into the problem by determining the optimal (that is, most likely) population trajectory leading to an unusually fast blowup.

It is instructive to compare the results of this work with those recently obtained \cite{KM2025} for the continuous model, described by the Langevin equation~(\ref{Langevin}), which shares the mean-field dynamics with the $2A\to 3A$ population model.

One crucial difference between Eq.~(\ref{Langevin}) and the $2A\to 3A$ model is that the continuous coordinate $x$ in Eq.~(\ref{Langevin}) can be negative, whereas the discrete number $n$ in the model $2A \to 3A$ is strictly positive. In spite of this difference, qualitative features of the blowup time distribution $\mathcal{P}(T)$ in the two models turn out to be quite similar. In particular, the long-time tail of $\mathcal{P}(T)$ for the  continuous model~(\ref{Langevin}) is also purely exponential, and the short-time tail also has an essential singularity at $T=0$. This essential singularity, however, is much stronger: $-\ln\mathcal{P}(T)\sim T^{-3}$ as compared with $T^{-1}$ in Eq.~(\ref{PsmallT}).

An important and somewhat surprising difference between the two models emerges at a formal level. When analyzing the blowup time statistics in the continuum model~(\ref{Langevin}), one encounters bi-confluent Heun functions \cite{KM2025}. These special functions are insufficiently studied, which complicates the analysis. The $2A \to 3A$ model proves to be significantly more tractable, as demonstrated by the relatively simple exact analytical results presented here for all quantities of our interest.

\bigskip
\noindent
{\bf Acknowledgments}. The author thanks M. Assaf, P. L. Krapivsky and N. R. Smith for useful discussions.  This research was supported by the Israel Science Foundation (Grant No. 1499/20).

\bibliography{2A3A}

\end{document}